\documentclass[
  ,draft            
  ]
  {aipproc}

\layoutstyle{6x9}

\begin{document}
\newcommand{\IP}{$\:I\!\!P\:$}
\newcommand{\DPE}{$\:DI\!\!PE\:$}
\begin{flushright}
FERMILAB-CONF-05-338-E
\end{flushright}
\title{Double Pomeron Physics at the LHC}


\author{Michael G. Albrow}{
  address={Fermi National Accelerator Laboratory, Batavia, IL 60510, USA}
}

\begin{abstract}
 I discuss central exclusive production, also known as Double Pomeron Exchange,\DPE , from the
 ISR through the Tevatron to the LHC. There I emphasize the interest of exclusive
 Higgs and $W^+W^-/ZZ$ production.
\end{abstract}

\maketitle


\section{Introduction}

     In 1973, shortly after the CERN Intersecting Storage Rings (ISR) provided the first
 colliding hadron beams, ``high mass" diffraction was discovered by the
 CERN- Holland- Lancaster- Manchester collaboration~\cite{chlm}. In this context ``high
 mass" meant $\approx$ 10 GeV, much larger than the $\approx$ 2 GeV diffractive states
 seen hitherto. Then Shankar~\cite{shankar}  and
 D.Chew and G.Chew~\cite{chew} predicted in
 the framework of Triple-Regge theory double pomeron exchange,\DPE,where
 both beam hadrons are coherently scattered and a central hadronic system is produced.
 Later experiments, in particular at the Split Field Magnet~\cite{sfmdpe} and the Axial Field
 Spectrometer (AFS)~\cite{afsdpe} discovered the processes:
 \IP\IP $\rightarrow \pi^+\pi^-,K^+K^-,p\bar{p},4\pi$ at $\sqrt{s}$ up to 63 GeV. In the case of
 the AFS we added very forward proton detectors to the large central high-$p_T$ detector,
 motivated largely by a search for glueballs. Structures were indeed found in the $\pi^+\pi^-$
 mass spectrum, not all understood and not, unfortunately, studied at higher $\sqrt{s}$. The
 absence of a $\rho$ signal verified that \DPE is indeed dominant at this energy, but not at
 lower (SPS) energies. Measuring the (coherently scattered) forward protons allowed a partial wave analysis to select
 $J=0,2$ central states.
 
    Now we want to do a similar experiment on a much grander scale, adding very small forward
 proton detectors to the large central high-$p_T$ detectors: CMS and ATLAS. At $\sqrt{s}$ =
 14,000 GeV rather than 63 GeV we will be measuring $W^+W^-$ and $ZZ$ rather than $\pi^+\pi^-$
 and looking for Higgs bosons or other phenomena (perhaps even more interesting, such as
 anomalous EWK-QCD couplings). What will the $M(W^+W^-), M(ZZ)$ spectra look like? As
 at the ISR, measurements of the (coherently scattered) forward protons will enable one to determine the quantum numbers of
 the central states, picking out the S-wave (scalars), D-wave (spin 2), etc. This is very powerful;
 even if, for example, a Higgs boson is discovered another way it may take central exclusive
 production to prove that it is a scalar. There will be forward roman pots around CMS at
 220m for the
 TOTEM experiment to measure (in special runs) $\sigma_{TOT}, \frac{d\sigma}{dt}$ and other
 diffractive processes. To study central masses below 200 GeV (the favored Higgs region) in
 normal high luminosity low-$\beta$ running we need to measure protons even farther from the
 collision point, at 420m. Physicists from ATLAS, CMS and TOTEM have joined forces on an R\&D
 project called FP420 to develop common technical solutions; we hope both large detectors will have this
 proton tagging capability.
 
    In symmetric colliding beams the beam rapidity $y_{BEAM} = ln
 \frac{\sqrt{s}}{M_p}$, and a central produced state of mass $M_{CEN}$ spans
 approximately $\Delta y_{CEN} = 2 ln \frac{M_{CEN}}{M_0}; M_0 \approx 1$ GeV. Pomeron \IP exchanges begin to
 dominate (exceeding Reggeon exchanges) when a rapidity gap exceeds about 3 units, which is a
 good ``rule-of-thumb", although 4 units is safer. Requiring two gaps of $>$ 3
 units, the maximum central mass follows from the above as simply $M_{CEN}(max)
 \approx \frac{\sqrt{s}}{20}$, which gives nominal limits of 3 GeV at the ISR
 (less at the SPS fixed target, which is therefore very marginal), 100 GeV at
 the Tevatron and 700 GeV at the LHC. The central exclusive mass spectra
 did indeed extend to $\approx$~3 GeV at the ISR~\cite{afsdpe}, and the 
 Tevatron experiment CDF finds~\cite{cdfdpe} 
 \DPE di-jets with masses up to $\approx$ 100 GeV. The Tevatron would be a
 perfect place for low mass \DPE spectroscopy (glueballs, hybrids, odderon
 search) but this has not yet been done. At the $Sp\bar{p}S$ collider, with
 $\sqrt{s}$ = 630 GeV, a few \DPE studies were done. UA1 had no forward proton
 detection but studied~\cite{ua1dpe} charged multiplicity $n^\pm$ and $p_T$ distributions up to $M_{CEN}\approx$ 60
 GeV using rapidity gaps. UA8 had roman pots, but studied mostly single diffraction, with some
 low mass \DPE \cite{ua8dpe}. At the Tevatron ($\sqrt{s}$ = 630, 1800, 1960 GeV) CDF has forward
 proton (FP) detection (roman pots) on the $\bar{p}$ side only, and uses the gap criterion on
 the $p$ side. As well as jet physics, searches are underway for exclusive
 $\chi_c$ and exclusive central $\gamma\gamma$ without, unfortunately, detecting
 the protons. D0 in Run~1 had no FP detection but studied jets with gaps. In Run~2
 they now have FP detection on both sides but have not presented \DPE data
 yet. 
 
 The extension of the \DPE mass range
 from $\approx$100 GeV at the Tevatron to $\approx$700 GeV at the LHC is
 exciting, as it takes us into the $W,Z,H,t\bar{t}$ domain.
  
\section{Central Exclusive Production at the LHC}

    The main channel for Higgs boson production at the LHC is $gg$-fusion. Another gluon exchange
can cancel the color and can even leave the protons intact: $pp\rightarrow p+H+p$ where the $+$
denote large rapidity gaps and there are \emph{no} other particles produced (i.e. it is \emph{exclusive}). If the outgoing
protons are well measured, the mass $M_{CEN} = M_H$ can be determined by the missing mass
method~\cite{albmm} with $\sigma_M \approx$ 2 GeV, and its quantum numbers can be determined. Theoretical
uncertainties in the cross section involve skewed gluon distributions, gluon $k_T$, gluon
radiation, Sudakov form factors, etc. Probably~\cite{kmr1,kmr2} for a Standard Model (SM) Higgs,
$\sigma_{SMH} \approx$ 0.2 fb at the Tevatron, which is not detectable, but at the LHC
$\sigma_{SMH} \approx$ 3 fb (within a factor 2-3) and with the higher luminosity (30-100
fb$^{-1}$) there should be enough events to be valuable. Some of the uncertainties in the cross
section can be addressed by measuring related processes at the Tevatron. The process
$gg\rightarrow H$ proceeds through a top loop. The same diagram with instead a $b(c,u)$ loop can
give exclusive $\chi_b(\chi_c,\gamma\gamma)$, which can therefore be used to ``calibrate" the theory
now at the Tevatron and then in the early days of the LHC. There are predictions~\cite{kmr3,kmr1} for exclusive
$pp\rightarrow p+ \chi_c+p \approx$ 600 nb at the Tevatron, $\approx$ 20/sec! In reality requiring decay
to a useful channel ($\chi_c\rightarrow J/\psi\gamma\rightarrow \mu^+\mu^-\gamma$), no other interaction
(for cleanliness), trigger efficiency and acceptance reduces this to effectively a few pb (still, thousands of
events in 1
fb$^{-1}$). Candidates have been seen (also candidates for exclusive $J/\psi$ which may be from
photoproduction ($\gamma$\IP)). Exclusive $\chi_b$ may also be possible but is marginal; the cross section is 5000 times
smaller. It will be valuable to measure this early at the LHC (with TOTEM+CMS at high $\beta^*$?). Unfortunately in CDF we cannot detect the associated protons, which would provide a quantum
number filter, selecting mainly $I^GJ^{PC}=0^+0^{++}$; $J^P=2^+$ is forbidden at $t$ = 0 for a $q\bar{q}$ state. D0 \emph{may}, but they have
$|t_{min}| \approx$ 0.7 GeV$^2$ which limits the statistics.  

   The process $pp\rightarrow p+H+p$  with $H\rightarrow b\bar{b}$ with no other 
activity (e.g. no gluon emission) would have two and only two central jets. We can also have $pp\rightarrow
p+gg+p$ or $pp\rightarrow p+b \bar{b}+p$ which we call ``exclusive dijets", although it is clear that both experimentally and theoretically
that is not a well defined state (unlike exclusive $\chi_c$ or exclusive $W^+W^-$ production). Nonetheless
we look in CDF for signs of ``exclusive dijets" which we can define, with some arbitrariness, as events
where two central jets as defined by a jet algorithm (again, not unique) have
$R_{jj}=\frac{M_{jj}}{M_{CEN}}>0.8$. (The events selected have a forward $\bar{p}$ detected and a rapidity
gap on the $p$-side.) There is no $R_{jj}$ = 1 ``exclusive" peak, and probably none is expected; there may
be a broad high $R_{jj}$ enhancement but with respect to what? CDF look to see if at $R_{jj} >$ 0.8 there
is a depletion of quark (specifically $b$) jets as expected~\cite{kmr3}; we can also look at
the $g/q$-jet ratio using internal jet features vs $R_{jj}$. At the LHC, one could get very large samples
(early, with low luminosity, tagging the protons) of exclusive dijets with $M_{CEN} = M_{jj} \approx$ 100-200 GeV.
These should be very pure gluon jets, which could be used to study QCD (think of the large samples of
quark jets studied at LEP on the $Z$).

    A difficult issue with exclusive SMH(120-130 GeV) is that the 420m $p$-detectors are too far away to be
included in the 1st level trigger, L1, and the central jets from the $H$-decay are completely overwhelmed
by QCD jet production. Putting forward rapidity gaps in the L1 trigger can be done but only works with
single interactions/low luminosity. The total integrated luminosity if only single interactions can be
used is expected to be $\approx$ 2-3 fb$^{-1}$ which is not enough for a SM Higgs, although it might be
for some MSSM scenarios which can have a much bigger (factor $\approx$ 50) cross section. 
[J.Ellis, J.Lee and A.Pilaftsis discussed~\cite{jellis} diffractive production of MSSM Higgs at the LHC.] A solution
might be to have a L1 trigger based on a 220m pot track and 2 jets with specific kinematics, such as 100 GeV
$<M_{jj}<$150 GeV, small $\sum \vec{E_T}$ (the forward protons will have $\sum\vec{p_T}<\approx$ 2 GeV and the jets
balance that), and with the jets in the same rapidity hemisphere as the 220m proton. Better, for the desired process $pp\rightarrow p+J_1 J_2 + p$ there is a relation between the
rapidities of the jets $y_1,y_2$ and the momentum loss fractions $\xi_1,\xi_2$ of the forward protons:
$\xi_{1(2)} = \frac{E_T}{\sqrt{s}}[e^{-(+)y_1}+e^{-(+)y_2}]$. If a (even a few-bit) measurement of $\xi$ from
the 220m pot track can be combined with the jets' ($E_T,\eta,\phi$) at L1 it should help, but the technical
feasibility (and value) remains to be studied. The 420m detectors can be included at L2.
If the Higgs boson mass is 140 - 200 GeV $W^+W^-$ and eventually $ZZ$ decays come in, and can provide L1
triggers, so the forward detectors can be part of L2. They can again be of great value for quantum number
determinations and for a good mass measurement ($\sigma(M_H)$ = 2 GeV \emph{per event}) even in the mode
$W^+W^-\rightarrow l^+\nu l^- \bar{\nu}$. These events are clean even with pile-up, as the $l^+l^-$ vertex has no
other particles (the distribution of $n^\pm$ on the $l^+l^-$ vertex will be broad but with a peak at 0). Two
photon processes $\gamma\gamma\rightarrow W^+W^-$ give a continuum background for $WW$ (not for $ZZ$) and
the $|t|$ of the protons is smaller which
helps the rejection.

\section{Prompt Vector Boson Pair Production}

    By ``prompt" I mean not from $t\bar{t}$ (a most prolific source) and not from Higgs, and by ``vector boson" $V$ I mean
$\gamma, W, Z$. There are several production mechanisms.
Approximately 90\% of prompt $W^+W^-$ are from $q\bar{q}$ annihilation with $t$-channel $q$ 
exchange. This can produce any $Q=0,1$ pair. $q\bar{q}$ annihilation with an $s$-channel $V$ can produce only
$\gamma W, WW$ and $WZ$; it is important as a probe of the $VVV$ vertex. Virtual $V$ emission from quarks,
rescattering to a real pair (any pair, even $W^+W^+$) is negligible at the Tevatron, but is $\approx$ 10\% at the LHC
($WW$-scattering with a possible Higgs pole, something else, or unitarity violation!). Two photon production
$\gamma\gamma\rightarrow W^+W^-$ is about 100 fb at the LHC, is well known and has the characteristic feature of very
small $t_1,t_2$ for the protons. \DPE production of $W$-pairs has not been calculated, but whether inclusive or
exclusive it should be very small in the SM. I address later the possibility of this being dramatically wrong. Note
that in these various processes for $VV$ production there are different color flows (color triplet annihilation,
color singlet exchange, high $p_T$ forward jets, low-$p_T$ protons) which can give different hadronic activity. For
single interactions that might be of interest.

    At the Tevatron the (non-diffractive) cross sections agree with CTEQ NLO which predict: $\sigma(WW)$ = 12.4 pb, $\sigma(WZ)$ =
3.65 pb and $\sigma(ZZ)$ = 1.39 pb (the latter has not yet been measured). At the LHC the cross sections should be
$\approx 10\times$ higher. At the Tevatron we found the following ``rules-of-thumb" for diffractive production of hard final
states (jets, $W$): about 1\% (within a factor 2) are produced by single diffraction, and about $10^{-3}$
are produced by \DPE. This would imply 120 fb for SDE $\rightarrow W^+W^-$ and 12(1) fb for \DPE $\rightarrow
WW(ZZ)$ (+ anything). 

      The $WW$ decay mode of the SM Higgs rises through 10\% at 120 GeV, through
50\% at 140 GeV and is about 98\% above 160 GeV. Let us consider three $WW$ event classes at the LHC. In all cases consider only the $e\nu$ and $\mu\nu$
 decay modes, which unfortunately gives a factor ($4\times 0.106^2)$ = 0.045 (later we will relax this).
 The \DPE $WW\rightarrow l^+l^{'-}\nu\bar{\nu}+X$ cross section is $\approx$ 0.5 fb, small but perhaps not
 impossible to see; in any case this might be considered a background to the following more interesting signals.
 Exclusive $W^+W^-$ with the two forward protons and nothing else can come from exclusive Higgs production or
 from $\gamma\gamma$ collisions. The former is predicted to be, for a 170 GeV Higgs, $\approx$ 3 fb $\times$
 0.045 (BR) $\approx 0.13$ fb. The latter is larger, $\approx$ 100 fb $\times$ 0.045 = 4.5 fb. \emph{However}
 (a) the $\gamma\gamma$ data is a mass continuum while the Higgs events are localised with the missing mass method in a $\approx$ 4 GeV bin
 (b) the $t_1$ and $t_2$ of the protons is more peaked at low values in the
 $\gamma\gamma$ case. 
For both classes of exclusive events, with the $pWWp$ missing
mass method one can probably use
also the $\tau\nu$ decay mode and even the dominant $W\rightarrow q\bar{q}$
decay mode for one of the $W$'s. Note that there are potentially useful missing
mass games one can play, e.g. in $p_1p_2\rightarrow p_3+WW+p_4 \rightarrow p_3+l^\pm\nu j_1j_2+p_4$
the missing mass squared: $MM^2 = (p_1+p_2-p_3-p_4-p_e-p_{j_1}-p_{j_2})^2
=M_{\nu}^2 =0$. Ability to use $W\rightarrow q\bar{q}$ modes for one $W$ would increase the
statistics by a factor 7.4 over $e\nu,\mu\nu$ only. In $H\rightarrow ZZ \rightarrow \mu^+\mu^-\nu\bar{\nu}$ the
invisible missing mass from $MM^2 = (p_1+p_2-p_3-p_4-p_{\mu_1}-p_{\mu_2})^2 = M_Z^2$ should help distinguish this
from the $WW\rightarrow \mu^+\mu^-\nu\bar{\nu}$ state (of course we also have $(p_{\mu_1}+p_{\mu_2})^2 =
M_Z^2$), as well as measuring $M(ZZ)$.

We cannot expect to see \DPE $\rightarrow W^+W^-$ at the Tevatron, but it may 
still be very interesting to study the associated hadronic activity in $VV$ and also
single $V$ events. CDF and D0 each have around 20 $WW/WZ/ZZ$ events in Run 2 based on the first 0.2 fb$^{-1}$, 
with a factor $\approx$25 more to come. Counting associated hadrons in the CDF events we find a very large spread,
with $n_{ass}^\pm$ in $p_T > 400$ MeV/c, $|\eta|<1.0$ ranging from 0 to 34! More statistics and more studies are
needed to say if there is anything anomalous, and the ``super-clean" event cannot be called diffractive, but it is
likely that the high $n_{ass}^\pm$ event was a small impact parameter collision and the super-clean event had large
impact parameter \emph{and yet produced a $W$-pair}. 

\section{Different Pomerons}

To 0$^{th}$ order soft (low $|t|$, low $Q^2$) diffractive interactions are due to a pair of gluons in a color singlet
... a classical Low-Nussinov soft pomeron. There can be a small ($ggg$) component which becomes relatively 
more important at larger $|t|$. These exchanges are equivalent to a sum over towers of virtual glueballs. As $Q^2$
increases, $q\bar{q}$ evolve in. Reggeons are predominantly towers of virtual $q\bar{q}$ mesons, summed over spins.
There has been an ambitious attempt to calculate the pomeron in QCD as a ``reggeized gluon ladder" ... the BFKL
pomeron. It is known that the exchange of a single gluon between quark lines, the leading order $qq$-scattering
QCD diagram, is `sick"; it is not gauge invariant. A summing procedure over diagrams can result in a gauge invariant
exchange, the ``reggeized gluon". In the BFKL pomeron two reggeized gluons cancel each other's color. This
``pomeron exchange between quarks" diagram enhances jet production in the forward direction (low $|t|$, high $s$). In
the ``White pomeron"~\cite{white} the color of the reggeized gluon is cancelled instead by an infinite number of wee gluons (they
have no momentum even in the infinite momentum frame). The wee gluons have the properties of the vacuum; in a sense
they \emph{are} the vacuum. In White's theory asymptotic freedom requires a pair of very heavy color sextet quarks,
which couple strongly to the pomeron \emph{and} to the $W$ and $Z$ once the energy is high enough. Consequently at
the LHC diffractive $W,Z$ production should be prolific, including $pp\rightarrow p+WW/ZZ+p$ exclusive
states.There should also be an \emph{effective} $\gamma Z$\IP coupling through color sextet quark loops, and hence
photoproduction of single $Z$ seen as $pp\rightarrow p+Z+p$, which would be another surprise (effectively
an anomalous EWK-QCD coupling).

\section{FP420}
 
The potentially rich physics program at the LHC with \DPE , especially with central states $WW,ZZ,H,jj,t\bar{t},X$,
needs the big central detectors CMS and ATLAS together with very forward proton detection and precision measurement.
This can be partially provided by the TOTEM detectors with CMS, but it is necessary to supplement them with detectors
at 420m. At this place the relevant protons have been deflected out of the beam by $\approx$ 3-25 mm where they can be
detected in small precision pixel tracking detectors. An international consortium of CMS, TOTEM and ATLAS physicists
has been formed to develop this proposal, and a LOI for support for R\&D will be sent to the LHCC in
June.

The proposed precise very forward proton detectors may have a side benefit of
calibrating the energy scale of the hadronic calorimetry. (At the Tevatron this
gives the largest uncertainty in e.g. the top quark mass.) During a special
run at low luminosity with less than one interaction per crossing, trigger on
events with two forward protons and nothing else beyond (say) $\eta$ = 4.0
($\theta = 2^\circ$). The total central mass (e.g. $\approx$ 200 GeV) is contained in the main CMS/ATLAS
detctors and is known to $\approx$ 1\%. The electromagnetic calorimetry should
already be well calibrated with $Z\rightarrow e^+e^-$, so this calibrates the hadronic
energy scale, for jet or non-jet events. This should be competitive with other
approaches ($\gamma$-jet balancing and $W\rightarrow$ jet+jet in top events).

\section{Closing Remarks}

There are many other related talks at this meeting (e.g. Cox, Eggert, Klein, Kowalski, Piotrzowski, Royon, ...)
demonstrating the interest in the field. This is sure to be a very exciting field at the LHC, \emph{whether or not
the Higgs boson is in reach}. It it exists and we see it, central exclusive production will be important for measurements of the mass, quantum
numbers, couplings and other properties. If it does not exist, exotic new physics may manifest itself through this process.
We will have come a long way from $pp\rightarrow p+\pi^+\pi^-+p$ at
$\sqrt{s} = 63$ GeV to $pp\rightarrow p+W^+W^-+p$ at $\sqrt{s}$ = 14,000 GeV!

\end{document}